\documentclass[
10pt,A4paper,twocolumn,aps,pra, superscriptaddress,longbibliography,nofootinbib,amsmath,amssymb,
]{revtex4-2}

\usepackage[utf8]{inputenc} % utf8 input
\usepackage[T1]{fontenc}    
\usepackage{lmodern}        
\usepackage{microtype} % improved kerning
\usepackage{graphicx} % graphics
\usepackage{enumerate}
\usepackage[a4paper, total={7in, 9.5in}]{geometry}
\usepackage{amsmath}
\usepackage{amsfonts}
\usepackage{amssymb}
\usepackage{bm} % bold math
\usepackage{braket}
\usepackage{mathtools}
\usepackage{enumitem}
\usepackage[colorlinks=true, linkcolor=blue, citecolor=blue, urlcolor=blue, pdfauthor={M. Kerschbaum, et. al.}, pdftitle={Investigating the Sensitivity of Niobium- and Tantalum-Based Superconducting Qubits to Infrared Radiation}]{hyperref}
\usepackage[capitalize]{cleveref}

\usepackage{tabularx}
\usepackage{cellspace}
\usepackage{collcell}
\usepackage{booktabs}
\usepackage{newfloat}
\usepackage{caption}
\usepackage{siunitx}
\usepackage[dvipsnames]{xcolor}
\usepackage{chemformula}
\usepackage{xprintlen}
\usepackage{lipsum}
\usepackage{fancyhdr}
\usepackage{listings}
\usepackage{subcaption}
\usepackage{comment}
\usepackage{textgreek}

\usetikzlibrary{calc}

\DeclareSIUnit\permille{\text{\textperthousand}}

\crefformat{subequations}{Eqs.~(#2#1#3)}
\Crefformat{subequations}{Eqs.~(#2#1#3)}

\let\originalleft\left
\let\originalright\right
\renewcommand{\left}{\mathopen{}\mathclose\bgroup\originalleft}
\renewcommand{\right}{\aftergroup\egroup\originalright}

\newcolumntype{Y}{>{\centering\arraybackslash}X}
\newcolumntype{d}{c}
\newcolumntype{a}{c}

\let\c\lstinline

\begin{document}

\begin{abstract}
The use of tantalum films for superconducting qubits has recently extended qubit coherence times significantly, primarily due to reduced dielectric losses at the metal-air interface. 
However, the choice of base material also influences the sensitivity to quasiparticle-induced decoherence. 
In this study, we investigate quasiparticle tunneling rates in niobium and tantalum-based offset-charge-sensitive qubits. 
Using a source of thermal radiation, we characterize the sensitivity of either material to infrared radiation and explore the impact of the infrared background through the targeted use of in-line filters in the wiring and ambient infrared absorbers. 
We identify both radiation channels as significant contributions to decoherence for tantalum but not for niobium qubits and achieve tunneling rates of 100\,Hz and 300\,Hz for niobium and tantalum respectively upon installation of infrared filters.
Additionally, we find a time-dependence in the observed tunneling rates on the scale of days, which we interpret as evidence of slowly cooling, thermally radiating components in the experimental setup. 
% We discuss the observed differences between niobium and tantalum qubits in the context of quasiparticle diffusion. 
Our findings indicate that continued improvements in coherence times may require renewed attention to radiative backgrounds and experimental setup design, especially when introducing new material platforms.
\end{abstract}

\date{\today}

\author{Michael~Kerschbaum}
\thanks{These authors contributed equally to this work.}
\author{Felix~Wagner}
\thanks{These authors contributed equally to this work.}
\affiliation{Department of Physics, ETH Zurich, 8093 Zurich, Switzerland}
\affiliation{ETH Zurich - PSI Quantum Computing Hub, Paul Scherrer Institute, 5232 Villigen, Switzerland}
\affiliation{Quantum Center, ETH Zurich, 8093 Zurich, Switzerland}

\author{Uro\v{s}~Ognjanovi\'{c}}
\author{Giovanni~Vio}
\author{Kuno~Knapp}
\affiliation{Department of Physics, ETH Zurich, 8093 Zurich, Switzerland}
\affiliation{ETH Zurich - PSI Quantum Computing Hub, Paul Scherrer Institute, 5232 Villigen, Switzerland}
\affiliation{Quantum Center, ETH Zurich, 8093 Zurich, Switzerland}

\author{Dante~Colao~Zanuz}
\affiliation{Department of Physics, ETH Zurich, 8093 Zurich, Switzerland}
\affiliation{Quantum Center, ETH Zurich, 8093 Zurich, Switzerland}

\author{Alexander~Flasby}
\author{Mohsen~Bahrami~Panah}
\author{Andreas~Wallraff}
\author{Jean-Claude~Besse}
\affiliation{Department of Physics, ETH Zurich, 8093 Zurich, Switzerland}
\affiliation{ETH Zurich - PSI Quantum Computing Hub, Paul Scherrer Institute, 5232 Villigen, Switzerland}
\affiliation{Quantum Center, ETH Zurich, 8093 Zurich, Switzerland}

\title{Assessing the Sensitivity of Niobium- and Tantalum-Based Superconducting Qubits to Infrared Radiation}

\maketitle

Improvements in the coherence times of superconducting qubits have enabled increasingly complex experiments, scaling up both the number of qubits used and the number of gates executed~\cite{Acharya2025, Kim2023b}. 
These increases in circuit width and depth are primarily driven by advances in qubit design, experimental setups, control techniques, and fabrication-related improvements in coherence. 
In the context of qubit lifetimes, the use of tantalum as base layer material has shown promising results~\cite{Place2021}. 
However, while this material exhibits beneficial properties, such as a thin oxide layer~\cite{McLellan2023} with small loss tangent~\cite{Crowley2023}, it may also possess qualities that contribute to increased decoherence.
One notable contribution to decoherence arises from quasiparticles, which are generated when energy is deposited in the superconductor and Cooper pairs are broken. 
These quasiparticles can tunnel across the Josephson junction~--~a process that has been identified as a source of both energy relaxation and dephasing in superconducting qubits, as reported in early studies~\cite{Lenander2011, Catelani2012}.

No thermally excited quasiparticles are expected in the leads of the aluminum junctions at typical operation temperatures of less than 20\,mK and would noticeably appear only above 150\,mK~\cite{Diamond2022}. 
However, any form of energy deposition in the superconductor exceeding twice the superconducting gap can break Cooper pairs and produce non-thermal quasiparticles.
Proposed mechanisms for the generation of excess quasiparticles include spurious antenna modes of the superconducting circuit coupling ambient infrared radiation to the junction~\cite{Liu2024c, Rafferty2021, Pan2022}, pulse-tube vibrations~\cite{Kono2023}, relaxation of shear stress~\cite{Anthony-Petersen2024}, ionizing radiation from radioactive materials~\cite{Vepslinen2020, Cardani2023}, and impact of high-energy cosmic rays~\cite{Cardani2021, Loer2024}.
These high-energy events were observed to generate quasiparticles that propagate across the entire quantum device, causing correlated errors~\cite{McEwen2021b, McEwen2024}. 
Such correlated errors, if not mitigated, undermine the effectiveness of quantum error correction~\cite{Martinis2021a} and may present an obstacle to the scalability of quantum computing with superconducting qubits~\cite{Acharya2023, Acharya2025}. 
This challenge was addressed by the engineering of on-chip normal-metal quasiparticle and phonon traps~\cite{Riwar2016, Iaia2022} and variations in the film thickness and superconducting gap energy~\cite{McEwen2024, Kamenov2023}, reducing quasiparticle diffusion to the leads of and across the junction. 

\begin{figure*}[t!]
\includegraphics[width=1\textwidth]{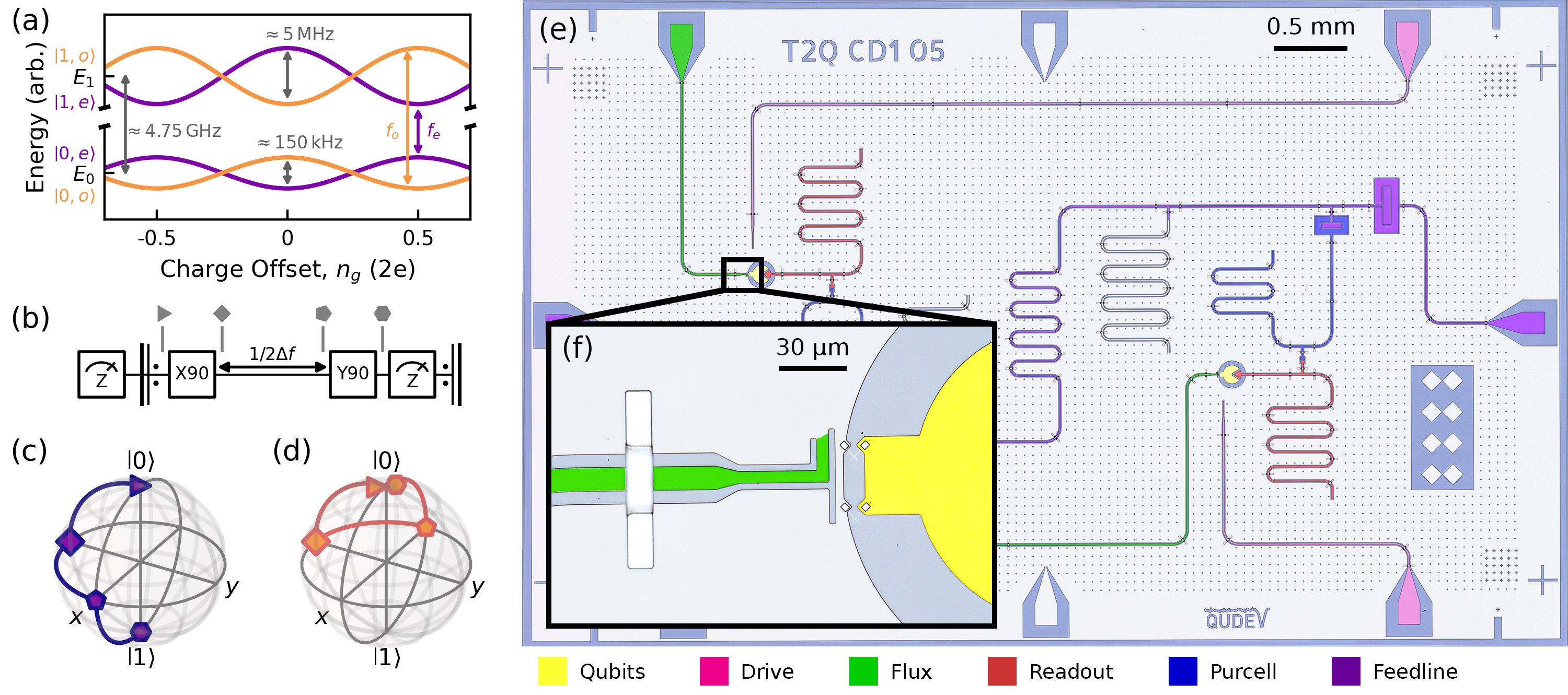}
\centering
\caption{Experimental protocol for quasiparticle tunneling measurement and design of the quantum device. (a)~Numerically obtained energy levels of an offset-charge sensitive transmon for typically observed device parameters. The eigenenergies of the transmon Hamiltonian depend on the offset charge $n_g$ on the qubit capacitor and show a periodicity of $2e$. (b) Experimental sequence for mapping the charge parity state of the qubit into the $\ket{0}/\ket{1}$ subspace. (c, d) Evolution of the qubit state on the Bloch sphere in case of (c) even and (d) odd charge parity when starting in the state $\ket{0}$. (e) False colored micrograph of one of the investigated quantum devices and (f) a zoom-in on the SQUID of one of the qubits, see main text for details.}
\label{fig:device}
\end{figure*}

In this work, we study quasiparticle generation in niobium and tantalum qubits with Al/AlO$_x$/Al junctions, focusing on how the choice of base material influences tunneling rates and qubit coherence times.
Using a Ramsey-type sequence~\cite{Riste2012a}, the charge parity state of a qubit is mapped onto the computational basis states, from which we extract the tunneling rate.
We assess the influence of out-of-equilibrium infrared radiation generated by a current biased resistive source operated near the quantum device on quasiparticle tunneling rates and coherence times. 
Using an experimental setup that follows the cryogenic engineering practices for superconducting quantum devices introduced in Ref.~\cite{Krinner2019}, we find quasiparticle tunneling rates to be higher in tantalum than in niobium, resulting in reduced coherence times for tantalum qubits.
However, by employing both in-line filters integrated into the coaxial control wiring and ambient infrared filtering using foam absorbers, we demonstrate how the performance of tantalum qubits can be improved such that they are no longer limited by quasiparticle tunneling-induced decoherence.
Furthermore, by tracking the tunneling rates over time after cooldown, we observe a reduction of the tunneling rates on the time scale of days and weeks, which we interpret to stem from gradually cooling components inside the cryostat.
We describe the measurement procedure in Sec.~\ref{sec:device}, analyze the effect of out-of-equilibrium infrared radiation on quasiparticle tunneling rates in Sec.~\ref{sec:ta_vs_nb}, present the impact of different infrared filtering strategies in Sec.~\ref{sec:filters}, and discuss the observed time dependence in Sec.~\ref{sec:decay}.

\section{Charge-sensitive qubits as sensors for quasiparticle tunneling}
\label{sec:device}

We resolve quasiparticle tunneling events by measuring offset-charge-sensitive transmons~\cite{Koch2007, Riste2012a} with a ratio of Josephson energy to charging energy $E_J/E_C\approx20$, corresponding to a charge dispersion of $\approx5\,$MHz for our qubits. 
The transmon eigenenergies are $2e$-periodic with respect to the charge offset $n_g$ on the qubit capacitor. 
For a given charge offset, the qubit will have one of two distinct frequencies, $f_e$ (even) and $f_o$ (odd), depending on the charge parity, as shown in Fig.~\ref{fig:device}a. 
Quasiparticle tunneling events change the charge parity and may consequently shift the qubit frequency. 
By resolving the qubit frequency at a rate faster than the typical tunneling rate, we can detect these jumps in charge parity and thereby resolve individual tunneling events.

We use a Ramsey-type sequence~\cite{Riste2012a} to extract the tunneling rates as detailed below. 
The qubit is initialized with a single-shot readout, heralding the $\ket{0}$ state, followed by a $\tau=50\,$ns long $X_{90}$ pulse, which excites the qubit to an equal superposition state on the equator of the Bloch sphere. 
For this operation, we drive the qubit at a frequency of $(f_e+f_o)/2$, corresponding to a frequency in between the ones of even and odd charge parity states. 
This is possible as the average Rabi-rate $\sim1/\tau=20\,$MHz is large compared to the detuning of the two frequency components of $\lesssim5\,$MHz. 
The qubit then evolves either clockwise or counterclockwise on the Bloch sphere in the frame of reference established by the first $X_{90}$ pulse, depending on its charge parity. 
By timing the subsequent $Y_{90}$ pulse at $1/(2|f_e-f_o|)=1/(2\Delta f)$ after the first pulse, we ensure that the two charge parity states are separated by a $\pi$ phase shift before being mapped onto the $\ket{0}$ and $\ket{1}$ states and read out, see Fig.~\ref{fig:device}b for the experimental sequence and Fig.~\ref{fig:device}c and d for the evolution of the qubit state on the Bloch sphere.

Instead of using active qubit reset as in Ref. \cite{Riste2012a}, our protocol employs a restless repetition of the measurement segment~\cite{Rol2017}. 
After each readout, we immediately continue with the next iteration, allowing for repetition rates of $0.5\,$-$\,1\,$MHz. 
% and thus enabling the measurement of quasiparticle tunneling rates of up to $200\,$kHz. 
% This implementation is possible as the protocol functions analogously when the qubit is initially in the excited state. 
% In this case, the qubit is mapped to the ground (excited) state for even (odd) charge parity. 
When analyzing the measurement traces, we distinguish between the qubit toggling between $\ket{0}$ and $\ket{1}$ for even charge parity and the qubit remaining in the previously measured state for odd parity. 
A quasiparticle tunneling event is then indicated by the qubit transitioning between the toggling and non-toggling state.

To measure quasiparticle tunneling rates we fabricated devices with target qubit frequencies of $5\,$GHz at their operation point. 
We perform single-qubit operations via a capacitively coupled drive line and operate the qubits~--~featuring an asymmetric superconducting quantum interference device (SQUID)~--~at their lower first-order flux-insensitive point~\cite{Blais2021}.
Single-shot readout of the qubits is achieved using pairs of readout resonators and Purcell filters, coupled to a shared feedline~\cite{Walter2017} and highlighted in red and blue, respectively, in Fig.~\ref{fig:device}e.

The niobium and tantalum devices are fabricated on a silicon substrate and employ Al/AlO$_x$/Al junctions (see App.~\ref{app:design_and_fab} for details on device design and fabrication). 
Each device is installed in its own nominally identical package and subsequently mounted into a multi-package sample holder, which accommodates up to three packages within the same magnetic shield assembly.
The experimental setup utilizes signal conditioning and routing as described in Ref.~\cite{Krinner2019}, with the addition of an in-line infrared filter in the readout output line after the parametric amplifier~(TWPA)~\cite{Macklin2015}. 
The impact of this additional in-line filter on tunneling rates is discussed in Sec.~\ref{sec:filters}.
Hence, our setup includes in-line infrared filters on all input and readout output lines and a mu-metal/aluminium/copper shield assembly to enclose the samples (see App.~\ref{app:setup} for information on the setup).

We first measure tunneling rates 
%~--~i.e., without operating the resistive infrared source~--~
for both niobium and tantalum devices in this configuration.
% The instantaneous qubit transition frequency in the even and odd parity state, $f_e$ and $f_o$, is determined by analyzing the beating pattern in an averaged Ramsey experiment. 
We execute the experimental sequence shown in Fig.~\ref{fig:device}b, taking a time trace of $\approx 500\,000$ integrated single-shot in-phase and quadrature pairs~\cite{Walter2017}, corresponding to about $1\,$s of measurement time. 
Reliable mapping of the charge parity state onto the qubit basis requires the charge offset to remain stable over the duration of the measurement~\cite{Riste2012a}. 
This condition is met in our experiments, as discussed in App.~\ref{app:procedure}.
% The qubit charge offset is stable on these timescales, see Ref.~\cite{Tennant2022} and App.~\ref{app:procedure}, ensuring a reliable extraction of the frequency components. 
We repeat the quasiparticle tunneling extraction experiment ten times, collecting datasets corresponding to a total of ten seconds of single-shot data. 
Additional details on the experimental protocol, along with a plot of the qubit frequency as a function of time, are provided in App.~\ref{app:procedure}.

\begin{figure}[!htbp]
\includegraphics[width=\linewidth]{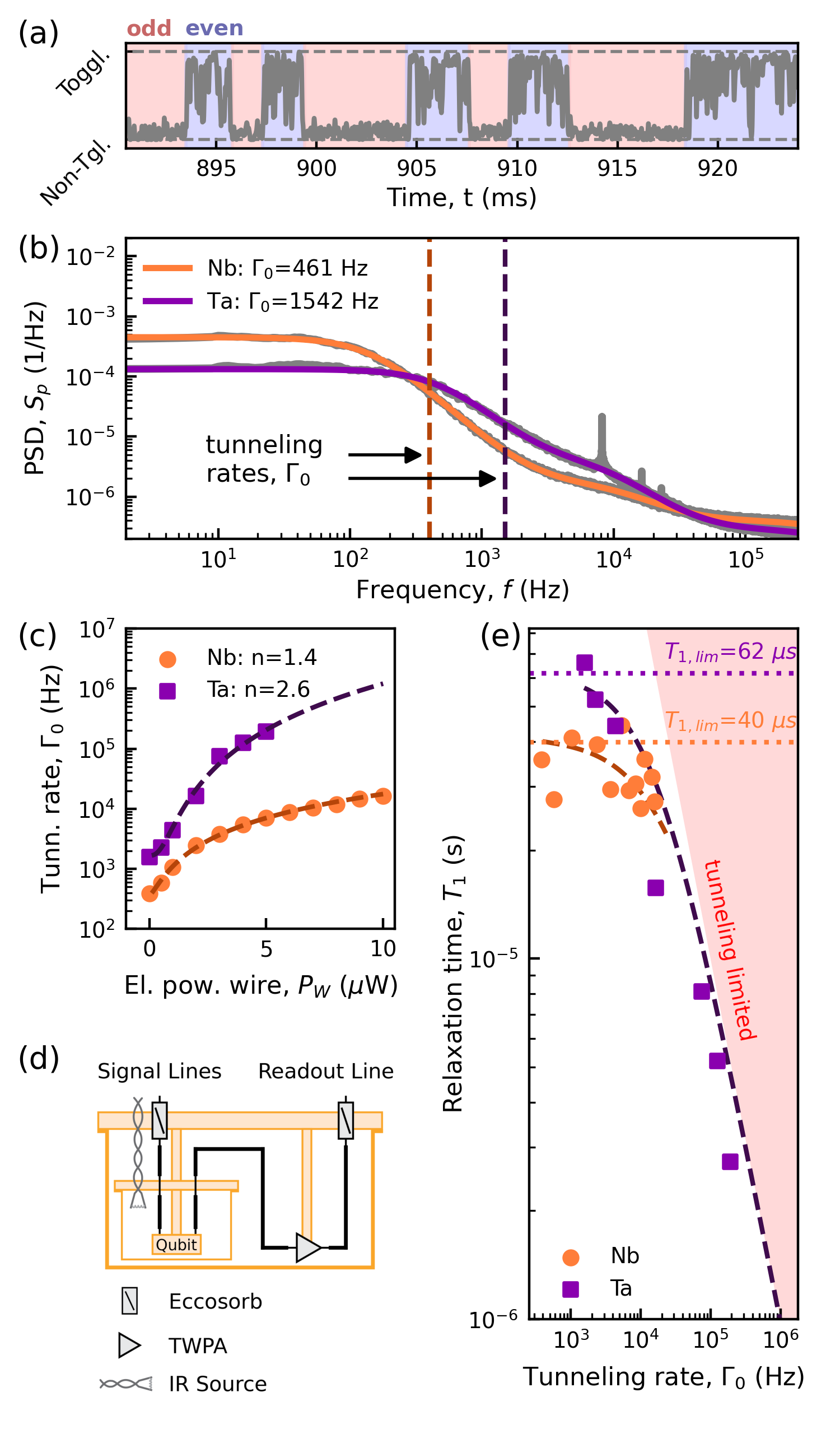}
\centering
\caption{Investigation of the sensitivity of niobium and tantalum devices to infrared radiation. 
(a)~Exemplary excerpt of a time-trace of the tunneling rate extraction experiment for a niobium qubit, with regions corresponding to even (odd) charge parity shaded in blue (red). 
The time trace shows the absolute value of the discrete derivative of the assigned qubit states (mapped to zero for $\ket{0}$ and one for $\ket{1}$), smoothed~--~for visualization purposes only~--~with a Gaussian filter of width $\sigma = 10$.
(b)~Example of a power spectral density dataset with applied moving average filter taken for both materials (gray dots), together with a Lorentzian model fit (solid lines, fitted to unfiltered data). 
(c)~Quasiparticle tunneling rate $\Gamma_0$ (taken for qubit one of device A (niobium) and device B (tantalum)) versus electrical power $P_W$ applied to the Manganin wire (dots), with power law fits (dashed lines) with fitted exponents~$n$. 
(d)~Schematic of the experimental setup. 
(e)~Correlation between tunneling rate and qubit relaxation time for data in (c). The region for which the qubit's $T_1$-time is limited by quasiparticle tunneling is indicated in red.
}
\label{fig:irsweep}
\end{figure}

This procedure yields a time series of toggling and non-toggling qubit state measurement outcomes through single-shot readout~\cite{Gambetta2007}, indicating the charge-parity state of the qubit as discussed above. 
We map the charge parity states to the values $1$ for even and $0$ for odd for further data processing, by computing the absolute difference of subsequent measurement outcomes of the assigned qubit states. 
An excerpt of a time trace for a niobium device, showing nine parity jumps in a time interval of $35\,$ms, is presented in Fig.~\ref{fig:irsweep}a. 
Subsequently, the absolute square of the discrete Fourier transform of the unfiltered time trace is computed to obtain the power spectral density. 
% The length of the time trace of each sequence is dependent on the wait time $1/2\Delta f = 1/2|f_e - f_o|$, which can vary between experimental runs, as it depends on the instantaneous value of the offset charge. 
% As a result, the power spectral density data is normalized for each measurement prior to averaging across all datasets.

The datasets of the power spectral density of parity time traces show a frequency dependence typical for random telegraph signals, with a flat plateau for low frequencies, indicating the absence of long-time correlations. Additionally, we observe a characteristic parity switching frequency above which the power spectral density decays as $1/f^2$, which we identify as the quasiparticle tunneling rate $\Gamma_0$, see Fig.~\ref{fig:irsweep}b. 
Notably, for this filter configuration we observe that the background tunneling rates in the tantalum qubits, at ($1542\,$ $\pm$ 176) Hz, are approximately three times higher than the ($461\,$ $\pm$ 16) Hz tunneling rate observed in the niobium qubits.
Further details regarding the fitting procedure including the effect of measurement infidelities are provided in App.~\ref{app:procedure}.
Finally, we note that the peaks observed at approximately $10\,$kHz are likely artifacts caused by electronic pickup, potentially resulting from non-ideal grounding of the experimental setup or insufficient filtering in the signal lines.

\section{Exposing Niobium and Tantalum devices to infrared radiation}
\label{sec:ta_vs_nb}

We generate infrared radiation inside the shield assembly by sourcing a mA-level current through a Manganin wire of $2\,$cm nominal length, $0.2\,$mm diameter and $2\,\Omega$ resistance~\cite{Diamond2022}, the location of which is indicated in the schematic of the experimental setup in Fig.~\ref{fig:irsweep}d. 
We measure the quasiparticle tunneling rates as a function of the electrical power dissipated in the Manganin wire, by varying the applied current. 
The rates exhibit a strong dependence on the dissipated power, spanning a dynamic range of up to two orders of magnitude. 
% and yielding maximal quasiparticle tunneling rates of $190\,$kHz for tantalum and $16.4\,$kHz for niobium.
% Additionally, tantalum devices not only exhibit a higher background rate, see Sec.~\ref{sec:device}, but also show a larger scaling of the tunneling rate as a function of the power of the induced infrared radiation. 
% A power law with a constant offset (corresponding to the base rate) is fitted to the datasets, yielding exponents of $n=1.34 \pm 0.04$ for niobium and $n=2.5 \pm 0.2$ for tantalum, see Fig.~\ref{fig:irsweep}c. 
Additionally, we fit power laws with a constant offset (representing the base rate) to the measured tunneling rates, yielding exponents of $n = 1.4 \pm 0.1$ for niobium and $n = 2.6 \pm 0.1$ for tantalum, as shown in Fig.~\ref{fig:irsweep}c.
These results indicate that, beyond exhibiting a higher background tunneling rate (see Sec.~\ref{sec:device}), the tantalum devices also show a stronger dependence of the tunneling rate on the radiated power, compared to niobium.
While the emission spectrum of the Manganin wire was not measured, thermal estimates suggest that the wire reaches temperatures of $\approx5\,$K at the highest applied power~\cite{Diamond2022}; we estimate the emitted thermal radiation to have a total power of up to 5\,nW.

We investigate correlations between the energy relaxation time $T_1$ and the quasiparticle tunneling rate. 
The relaxation rate is influenced by a combination of decay mechanisms, including quasiparticle-induced decay and factors unrelated to quasiparticle tunneling, such as the quality of the superconducting film interfaces. 
Tantalum qubits exhibit coherence times that appear to be limited by quasiparticle-tunneling-induced relaxation, even in the absence of externally applied infrared radiation.
This behavior is reflected in the non-zero slope of relaxation rate versus quasiparticle tunneling rate for tantalum qubits, even at the lowest applied electrical powers, see left-most data points in Fig.~\ref{fig:irsweep}e, indicating that increased quasiparticle tunneling directly impacts relaxation times.
In contrast, niobium qubits show no clear dependence of $T_1$ on the quasiparticle tunneling rate under comparable low applied power conditions.
These findings suggest that, in setups employing signal routing and conditioning as described in Ref.~\cite{Krinner2019}, the coherence times of tantalum-based qubits may primarily be limited by quasiparticle tunneling-induced relaxation, in contrast to niobium-based qubits, where film-quality-related factors dominate.

We observe significant differences in tunneling rates between devices using either tantalum or niobium as base-layer materials, despite all junctions being fabricated from Al/AlO$_x$/Al using an identical process. 
Furthermore, since the qubit designs are identical for both tantalum and niobium devices, qubit antenna modes~\cite{Rafferty2021, Liu2024c, Pan2022} coupling impinging infrared radiation directly to the tunnel junctions are unlikely to be the dominant factor in the differences between the observed rates.
Additionally, small qubit capacitor sizes, of diameter $\approx125\,$µm in our devices, have been shown to suppress the impact of antenna modes on quasiparticle tunneling rates~\cite{Pan2022}.

Our data suggest that the properties of the base-layer material play a crucial role in determining the quasiparticle tunneling rates in our devices.
One possible mechanism involves quasiparticles generated in the base layer diffusing toward the junctions, where they may tunnel across the barrier.
The effective quasiparticle flux toward the junction depends on their lifetime and diffusion constant, which determine the size of the region from which quasiparticles can reach the junction before recombining or being trapped~\cite{Catelani2011b}.
These quantities are sensitive to material composition, film thickness, fabrication process, and operating temperature.
Literature suggests, however, that quasiparticles in tantalum exhibit longer lifetimes and higher mobility than in niobium, potentially leading to an increased quasiparticle flux towards the junction in tantalum-based devices~\cite{Rando1992, Gijsbertsen1996, Yelton2024, Barends2008a, Nussbaumer2000}.

We hypothesize that, in our devices, a significant fraction of the observed tunneling events originates from such base-layer-generated quasiparticles.
Similar mechanisms have been proposed in prior work~\cite{Zhu2025a} and have been deliberately exploited to engineer quasiparticle traps~\cite{Riwar2016}.
Moreover, quasiparticle diffusion toward junctions or lower-gap superconductors forms the operating principle behind a range of quantum particle detectors~\cite{Angloher2016, Fink2021, Angloher2023, Fink2024, Ramanathan2024}.

% These findings suggest that a significant fraction of the observed tunneling rates may originate from quasiparticles initially created in the base-layer material, which subsequently diffuse toward the junction before tunneling across it. 
% Similar effects were reported in Ref.~\cite{Zhu2025a} and used in Ref.~\cite{Riwar2016} to engineer quasiparticle traps. 
% The diffusion of quasiparticles toward a junction or lower-gap superconductor is exploited in quantum devices for particle sensing~\cite{Angloher2016, Fink2021, Angloher2023, Fink2024, Ramanathan2024}.

% The quasiparticle lifetime and diffusion constant likely play a crucial role in the quasiparticle flux from the base material toward the aluminum tunnel junction. 
% The distance quasiparticles can diffuse before recombining or being trapped affects the size of the region around the junction leads that contributes to the quasiparticle influx. 
% These properties are highly dependent on factors such as material composition, fabrication process, film thickness, and operating temperature. 
% Literature suggests, however, that quasiparticles in tantalum are more mobile and have longer lifetimes compared to those in niobium, which may contribute to the observed differences in tunneling rates between the two materials~\cite{Rando1992, Gijsbertsen1996, Yelton2024, Barends2008a, Nussbaumer2000}.

The quasiparticle lifetime is limited by one of two termination mechanisms, trapping or recombination, as discussed in detail in Ref.~\cite{Catelani2011b}. 
In the first scenario, the equilibrium quasiparticle density~--~and thus the quasiparticle tunneling rate~--~is expected to scale quadratically with the power applied to the radiative source, whereas in a recombination-dominated regime, it is expected to scale linearly with the power (see App.~\ref{app:scaling} for a more detailed discussion). 
Our measurements of the tunneling rate scaling with the emitted power of the heating wire (Fig.~\ref{fig:irsweep}c) show a dependence of $\propto P_{W}^{1.34 \pm 0.04}$ for the niobium device and $\propto P_{W}^{2.5 \pm 0.2}$ for the tantalum device. 
While these exponents do not precisely match the idealized power laws (see App.~\ref{app:scaling}), the difference in scaling~--~approximately by a power of two~--~may indicate that recombination is the dominant quasiparticle termination mechanism in niobium, whereas trapping plays a more significant role in tantalum.

Additionally, differences in the superconducting gap energies of tantalum ($2\Delta_{\text{Ta}} / h \approx 340\,$GHz) and niobium ($2\Delta_{\text{Nb}} / h \approx 740\,$GHz)~\cite{Kittel2005} could influence the tunneling rate scaling. 
The higher gap in niobium results in suppressed photon absorption below this energy, which, in combination with a potentially power-dependent spectrum of our thermal radiation source, may contribute to the observed differences in the scaling of quasiparticle tunneling between niobium and tantalum devices.

\section{Reducing background tunneling rates through infrared filters} \label{sec:filters}

\begin{figure}[t!]
\includegraphics[width=\linewidth]{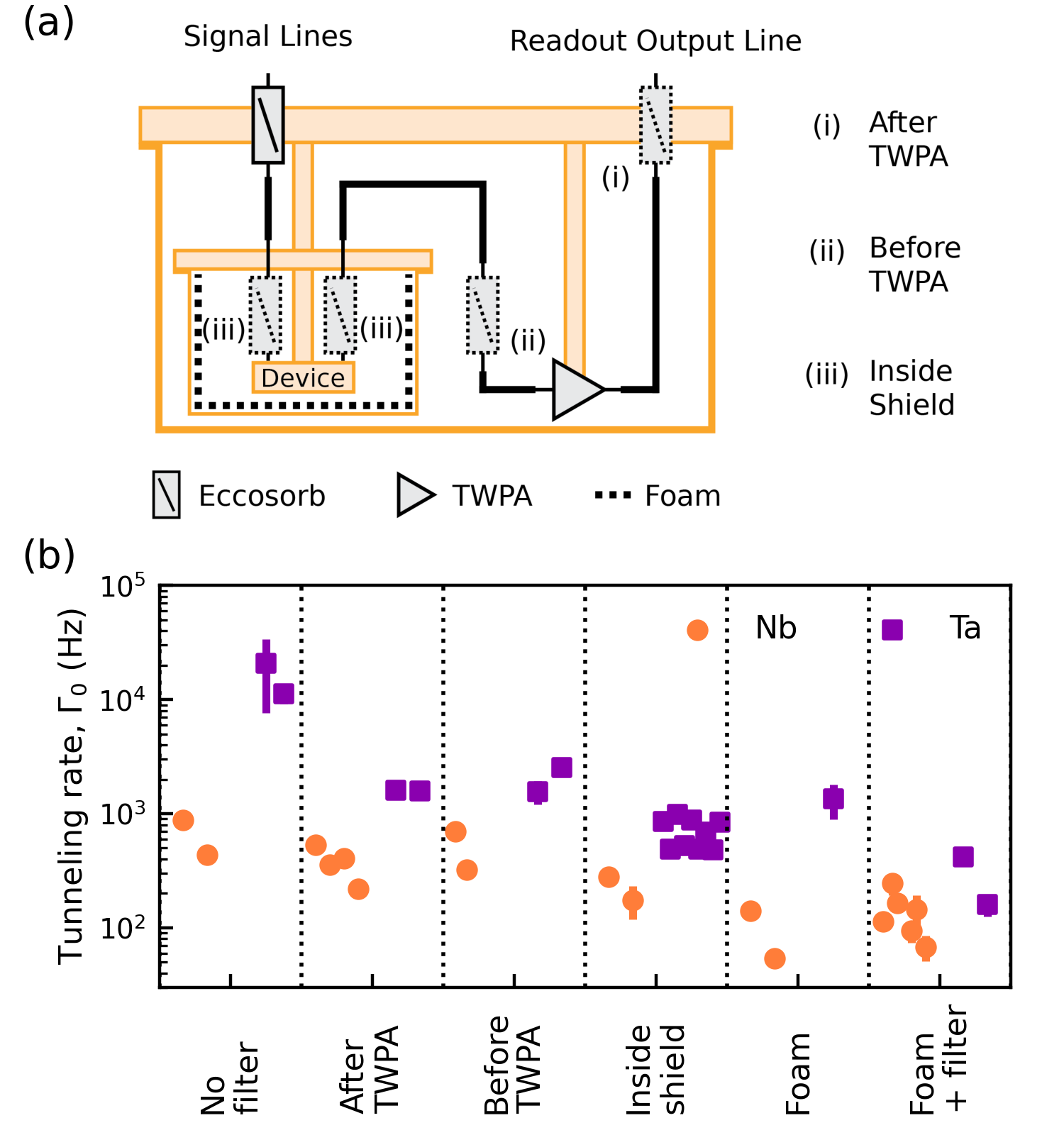}
\centering
\caption{Suppressing infrared radiative background through in-line filters and infrared-absorbing foam. (a)~Schematic of the experimental setup. Eccosorb filters are placed at three positions along the readout output and signal lines, before and after the TWPA. All filters are thermally anchored to the base plate using copper braids (not shown in the schematic). Eccosorb foam is placed inside the shield assembly. (b) Tunneling rates from multiple measurements and devices are compared for six filter and foam configurations, for niobium- and tantalum-based devices.}
\label{fig:filters}
\end{figure}

Having studied the effect of infrared radiation originating from a heated wire source on tantalum and niobium qubits, we employ infrared filters to mitigate background thermal radiation~\cite{Barends2011}.
For tests described in this section, we follow the signal conditioning scheme of Ref.~\cite{Krinner2019}, employing in-line infrared filters on charge and flux lines, but not on the readout output lines, unless noted otherwise.
Here, we follow two strategies to mitigate the infrared radiation background compared to this baseline configuration: first, we investigate the impact of in-line infrared filters in our readout lines, where we place them either after or before the traveling-wave parametric amplifier (TWPA), inside the shield assembly or not use any in the readout lines.
Secondly, we place a $\approx380\,$cm$^2$$\,\times\,1$\,cm layer of broadband infrared absorbing foam between the shield assembly and sample package. 
The setup schematics are shown in Fig.~\ref{fig:filters}a, details on the in-line filters, radiation-absorbing foam, light tightness of the sample package and shielding are discussed in App.~\ref{app:setup}.

\begin{table}[t!]
    \centering
    \begin{tabular}{l@{\hspace{0.5em}}l@{\hspace{1.5em}}r@{\hspace{0.3em}}c@{\hspace{0.3em}}l}
        \toprule
        Material & Configuration & \multicolumn{3}{c}{$\Gamma_0$ (Hz)} \\
        \midrule
        Nb & No filter        &  $10^2\times(6.60$  & $\pm$ & $2.29)$  \\
        Ta & No filter        &  $10^4\times(1.60$  & $\pm$ & $1.05)$  \\
        Nb & After TWPA       &  $10^2\times(3.79$  & $\pm$ & $1.13)$  \\
        Ta & After TWPA       &  $10^3\times(1.59$  & $\pm$ & $0.13)$  \\
        Nb & Before TWPA      &  $10^2\times(5.14$  & $\pm$ & $1.93)$  \\
        Ta & Before TWPA      &  $10^3\times(2.05$  & $\pm$ & $0.56)$  \\
        Nb & Inside shield    &  $10^2\times(2.27$  & $\pm$ & $0.67)$  \\
        Ta & Inside shield    &  $10^2\times(6.91$  & $\pm$ & $2.05)$  \\
        Nb & Foam             &  $10^2\times(0.98$  & $\pm$ & $0.45)$  \\
        Ta & Foam             &  $10^3\times(1.35$  & $\pm$ & $0.46)$  \\
        Nb & Foam + filter    &  $10^2\times(1.38$  & $\pm$ & $0.65)$  \\
        Ta & Foam + filter    &  $10^2\times(2.88$  & $\pm$ & $1.36)$  \\
        % Nb & No filter        &  $ 10^2\times(6.60$  & $\pm$ & $2.29)$  \\
        % Ta & No filter        & (1.60  & $\pm$ & $1.05)\times 10^4$ \\
        % Nb & After TWPA       & (3.79  & $\pm$ & $1.13)\times 10^2$ \\
        % Ta & After TWPA       & (1.59  & $\pm$ & $0.13)\times 10^3$ \\
        % Nb & Before TWPA      & (5.14  & $\pm$ & $1.93)\times 10^2$ \\
        % Ta & Before TWPA      & (2.05  & $\pm$ & $0.56)\times 10^3$ \\
        % Nb & Inside shield    & (2.27  & $\pm$ & $0.67)\times 10^2$ \\
        % Ta & Inside shield    & (6.91  & $\pm$ & $2.05)\times 10^2$ \\
        % Nb & Foam             & (0.98  & $\pm$ & $0.45)\times 10^2$ \\
        % Ta & Foam             & (1.35  & $\pm$ & $0.46)\times 10^3$ \\
        % Nb & Foam + filter    & (1.38  & $\pm$ & $0.65)\times 10^2$ \\
        % Ta & Foam + filter    & (2.88  & $\pm$ & $1.36)\times 10^2$ \\
        \bottomrule
    \end{tabular}
    \caption{Quasiparticle tunneling rates $\Gamma_0$ for different materials and filter configurations as plotted in Fig.~\ref{fig:filters}b.}
    \label{tab:configurations}
\end{table}

We observe a reduction in the quasiparticle tunneling rate with improved filtering, by comparing the effect of in-line filters placed at the three specified locations in the readout line. 
For tantalum, the mean tunneling rates are reduced by factors of 10.0, 7.8, and 23.2, respectively. 
For niobium, we observe reductions by factors of 1.7, 1.3, and 2.9 for the same positions.
While no significant difference in the rates is observed between placing the filter before or after the TWPA, positioning the filters inside the shield yields the strongest rate reduction.

Foam absorbers reduce the mean rate compared to the no-filter configuration by a factor 11.9 and 6.8 for tantalum and niobium respectively. 
Hence, for tantalum, both in-line filters and foam absorbers significantly reduce the quasiparticle tunneling rate, whereas for niobium, a reduction is primarily achieved by employing foam absorbers, with in-line filters having a smaller effect.
The combination of foam absorbers with in-line filters at either position (see table in App.~\ref{app:filter_data}) leads to the lowest observed quasiparticle tunneling rates for tantalum, with a reduction factor of 56. 
While the measurement protocol and qubit design differ, the observed quasiparticle tunneling rate for tantalum in this filter configuration agrees in order of magnitude with previously reported values~\cite{Tennant2022,Kurter2022}.
For niobium, however, the combination does not outperform the foam absorber configuration alone; within error bars, adding in-line filters to the setup with foam absorber does not further reduce the observed quasiparticle tunneling rates.
The measured data are plotted in Fig.~\ref{fig:filters}b, where individual points represent datasets of 50 averaged measurements. The averaged quasiparticle tunneling rates for all configurations are summarized in Tab.~\ref{tab:configurations}.
Notably, the lowest observed quasiparticle tunneling rates for both niobium and tantalum are several orders of magnitude lower than the inverse of the best observed coherence times in these devices. 
Since quasiparticle-induced decoherence occurs at a rate smaller than the quasiparticle tunneling rate, this suggests that coherence times in the millisecond range will not be limited by quasiparticle tunneling for the best filter configuration.

These results strongly suggest that the breaking of Cooper pairs by incident infrared radiation is one of the main origins of excess quasiparticles for devices in setups as described in Ref.~\cite{Krinner2019}.
Furthermore, since both in-line filters and foam absorbers independently affect the tunneling rate, we believe that infrared radiation propagates both in free space and in the cables leading to the devices. 
This was previously reported in Ref.~\cite{SerniakThesis}. 
The two mitigation methods are not completely independent, as their combined effect is smaller than the sum of their individual contributions. The in-line filters reduce tunneling rates by $14.4\,$kHz (Ta) and $0.28\,$kHz (Nb), while the foam absorbers reduce them by $14.6\,$kHz (Ta) and $0.56\,$kHz (Nb). 
However, when both are used together, the total reduction is only $15.7\,$kHz (Ta) and $0.52\,$kHz (Nb).
% This can be caused by a coupling mechanism between in-line and ambient infrared radiation, which is consistent with the measured higher effect of the in-line filters on mitigating the tunneling rates, when placed inside the shield assembly.
This effect may be explained by ambient infrared radiation entering the setup through insufficiently light-tight connectors or shielding. 
Such radiation can couple into the signal lines and propagate toward the quantum device, which would be consistent with the observation that in-line filters are most effective at suppressing quasiparticle tunneling rates when placed inside the shield assembly.

The observed qualitative differences in the effect of in-line filters on tunneling rates between niobium and tantalum could be explained by a potential difference in the frequency spectrum of in-line versus ambient radiation. 
Higher-frequency radiation originating from higher-temperature stages of the cryostat may play a dominant role for niobium due to its larger superconducting gap energy. 
Additionally, we did not observe a significant difference in tunneling rates between configurations where in-line filters were placed before or after the TWPA, suggesting that neither the amplifier itself nor the associated pump signal routing has a relevant impact on the radiation background at the observed tunneling rates.
\section{Observing Time-Dependent Tunneling Rates}\label{sec:decay}

\begin{figure}[t!]
\includegraphics[width=\linewidth]{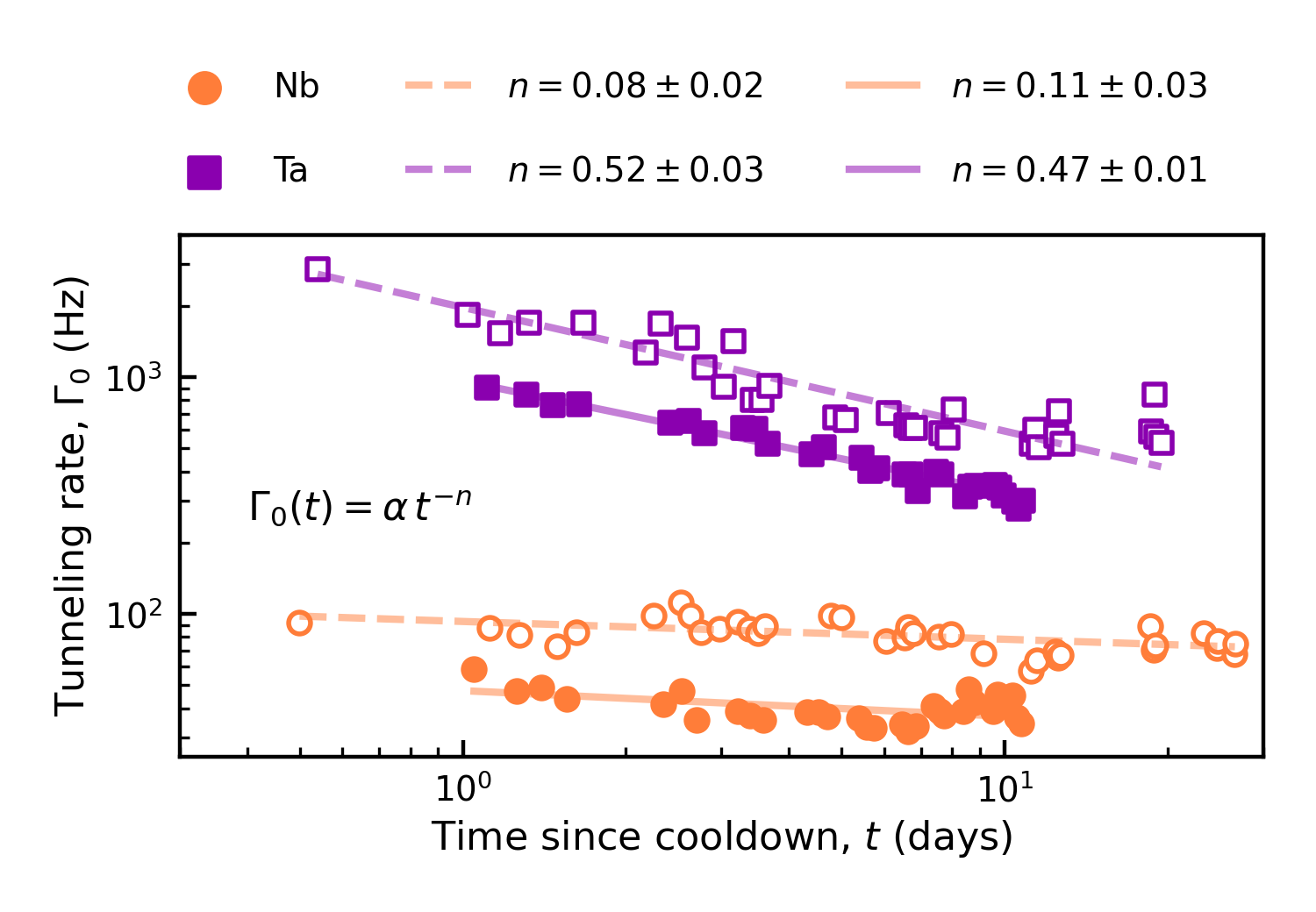}
\centering
\caption{Reduction in background quasiparticle tunneling rates over the course of a cooldown. The tunneling times for one niobium and one tantalum device are compared across two thermal cycles: one with Eccosorb foam installed (filled markers) and one without (empty markers). In all cases, we fit a material-characteristic power law to the data, see main text for details.}
\label{fig:decay}
\end{figure}

Daily measurements over several weeks reveal a systematic reduction in quasiparticle tunneling rates with time since start of the cooldown, an effect that is observed during multiple subsequent cooldown cycles. 
Plotting the tunneling rate as a function of time~$t$ since the start of the condensation process~--~defined as the moment when the base plate temperature first drops below 4 K, typically 52 to 57 hours after switching on the pulse tube coolers starting from room temperature~--~shows this reduction in the quasiparticle tunneling rates for both niobium and tantalum devices.
We measured the long-time changes in the quasiparticle tunneling rates for two thermal cycles: in the first, the setup included in-line filters in the readout output line before the TWPA (\textit{Before TWPA} in Fig.~\ref{fig:filters}b); in the second, we added foam absorbers inside the sample mount shield assembly (\textit{Foam + filter} in Fig.~\ref{fig:filters}b). 
Background quasiparticle tunneling rates were measured daily over more than three weeks in the first cycle (open markers in Fig.~\ref{fig:decay}) and two weeks in the second cycle (filled markers), for both niobium and tantalum devices.

We observe a clear power-law reduction of the quasiparticle tunneling rates in tantalum devices over time, and a weaker, yet similar, trend for niobium devices.
% We find the rate for the niobium device decreasing on a power-law time scale with a coefficient of $t^{-0.08\pm0.02}$ in the first run and $t^{-0.11\pm0.03}$ in the second run, where $t$ is measured in days. 
% The tantalum device exhibits a stronger time dependence, with coefficients of $t^{-0.52 \pm 0.03}$ in the first run and $t^{-0.47 \pm 0.01}$ in the second run.
% To confirm the presence of a time dependence~--~that is, a nonzero power-law exponent~--~we perform an F-test~\cite{Sachs2006}. 
% The significance of the time dependence in the niobium device is 3.3\,$\sigma$ in the first run and 2.6\,$\sigma$ in the second run, indicating a power-law trend, while for tantalum the power-law behavior is established with a significance exceeding 8\,$\sigma$ in both runs.
The improved filtering of ambient infrared radiation reduces the rate at $t=1$\,day from (93 $\pm$ 4)\,Hz to (48 $\pm$ 3)\,Hz for niobium, and from (1.97 $\pm$ 0.07)\,kHz to (0.96 $\pm$ 0.02)\,kHz for tantalum, in both cases by a factor two. 
Notably, the rate of decrease over time, however, remains unaffected by the additional foam absorbers.
The systematic decrease in quasiparticle tunneling rates with time contributes to the spread observed in Fig.~\ref{fig:filters}b, where rates for nominally identical configurations vary by up to a factor of $\approx2.5$, reflecting differing measurement times relative to the start of the cooldown.
Nevertheless, our conclusions regarding the effects of different filter configurations remain significant after accounting for this time dependence, as discussed in App.~\ref{app:filter_data}. 
A full list of data sets, including their uncertainties, measurement times, and associated devices and qubits, is also provided in~App.~\ref{app:filter_data}.

We attribute the observed reduction in quasiparticle tunneling rates over time since cooldown to thermal radiation from one or more components that are poorly thermally anchored and cool slowly on the timescale of weeks. 
Possible candidates include the aluminum shielding, which may thermalize poorly after becoming superconducting, and polymer-based dielectrics in coaxial cables and microwave components, exhibiting very low thermal conductivity~\cite{Drobizhev2017,NISTTeflon}.
A similar time-dependent decrease in quasiparticle tunneling rates was previously reported in Ref.~\cite{Mannila2021}, where the origin of the excess quasiparticle density remained unclear.
Comparable long-term relaxation phenomena have also been observed in superconducting dark matter and neutrino detectors~\cite{Adari2021, Leane2022, Angloher2023a}, where the authors report a background of unknown origin at low energies that decrease over time. 
Low-energy backgrounds were observed to stem from interfacial stress, contributing to excess quasiparticle densities in context of both cryogenic calorimeters~\cite{Anthony-Petersen2024} as well as qubit samples~\cite{Yelton2025}. 
However, interface stress is unlikely to be the dominant source of the tunneling rates observed in our experiments, as it would not be affected by the use of foam absorbers.
\section{Conclusion}\label{sec:disc}

In summary, we have demonstrated that the coherence time of tantalum qubits in typical superconducting qubit setups, such as the one described in Ref.~\cite{Krinner2019} can be limited by infrared-radiation-induced quasiparticles. 
We attribute the observed decrease in background quasiparticle tunneling rates over days and weeks after the beginning of the cooldown of the experimental setup to the slow cooling of thermally radiating components in our cryostat, which appear to be significant contributors to the infrared background. 
By implementing additional infrared filtering~--~both in the form of foam absorbers to tackle ambient infrared radiation and Eccosorb filters to tackle infrared radiation in the readout lines~--~we achieve a substantial reduction in tunneling rates, reaching levels that are negligible for qubit coherence times in the millisecond range.

Understanding low-energy radiative backgrounds is crucial for optimizing the performance of quantum devices. 
Our findings suggest that as superconducting qubit coherence times continue to improve or new material platforms are introduced~\cite{Place2021}, revisiting the configuration of experimental setups may become necessary~\cite{Krinner2019}. 
Additionally, beyond addressing antenna modes that contribute to quasiparticle tunneling via resonant absorption~\cite{Liu2024c}, a systematic investigation of quasiparticle diffusion and its role in quasiparticle poisoning at the junction may be essential for developing effective mitigation strategies.

\section*{Acknowledgements}
We thank Moritz Bürgi for his contributions to the measurement and analysis framework. This work was funded by the Swiss State Secretariat for Education, Research and Innovation (SERI).
Research was sponsored by IARPA and the Army Research Office, under the Entangled Logical Qubits program, and was accomplished under Cooperative Agreement Number W911NF-23-2-0212.
The authors acknowledge financial support by the Baugarten Foundation, the ETH Zurich Foundation, and by ETH Zurich.
The views and conclusions contained in this document are those of the authors and should not be interpreted as representing the official policies, either expressed or implied, of IARPA, the Army Research Office, or the U.S. Government. The U.S. Government is authorized to reproduce and distribute reprints for Government purposes notwithstanding any copyright notation herein.

\section*{Author Contribution}
M.K., F.W. and J.C.B. planned the experiments, with M.K., F.W., U.O., and G.V. taking measurement data. M.K. and F.W. analyzed the data with contributions from U.O, G.V. and K.K.
K.K. designed the device with input from J.C.B., and A.F., M.B.P. and J.C.B. fabricated the devices. 
M.K. designed and built elements of the experimental setup with contributions from F.W. and U.O. 
M.K., F.W., U.O. and G.V. characterized and calibrated the devices and the experimental setup. 
M.K. and F.W. prepared the figures for the manuscript, and 
M.K., F.W. wrote the manuscript with inputs from all co-authors. 
A.W. and J.C.B. supervised the work.

% \clearpage
\appendix
\section{Quantum Device Design and Fabrication}
\label{app:design_and_fab}

The devices used in this work feature two offset-charge-sensitive transmons each, with targeted charge dispersion of $\epsilon_{01}\approx1$-$5\,$MHz. Qubit state control is achieved via capacitively coupled charge lines. The Josephson junctions are implemented as superconducting quantum interference devices (SQUIDs), allowing tunability of the Josephson energy through direct current biasing via an inductively coupled flux line. Each qubit is read out via an individual readout resonator and Purcell filter pair, coupled to a common feedline. Values for relevant qubit parameters measured at their operational point are listed in Tab.~\ref{tab:target_device}.

\begin{table}[b!]
    \centering
    \begin{tabular}{l@{\hspace{1em}}l@{\hspace{1em}}rl}
        \toprule
        Parameter & Symbol & \multicolumn{2}{l}{Meas. Range} \\
        \midrule
        Qubit freq. & $\omega_\text{LSS}/2\pi$ & 4.80 - 5.55 & GHz\\
        Readout res. freq. & $\omega_\text{r}/2\pi$ & 7.44 - 7.68 & GHz\\
        Charge disp. & $\epsilon_{01\text{,LSS}}$ & 1.3 - 4.5 & MHz\\
        Anharmonicity & $-\alpha$ & 460 - 471 & MHz\\
        Qubit lifetime & $T_\text{1,LSS}$ & 30 - 80 & µs\\
        Qubit coherence time & $T^*_\text{2,LSS}$ & 5 - 20 & µs\\
        \bottomrule
    \end{tabular}
    \caption{Measured device parameter ranges for the offset-charge sensitive transmons provided at the operational point, i.e., the lower first-order flux-insensitive point (LSS). Qubit lifetimes as measured with infrared filter in readout output line, see App.~\ref{app:setup}.}
    \label{tab:target_device}
\end{table}

\begin{table}[t!]
    \centering
    \begin{tabular}{l@{\hspace{1em}}l@{\hspace{1em}}rl}
        \toprule
        Design Parameter & Symbol & \multicolumn{2}{c}{Value} \\
        \midrule
        Lead length & $l_\text{lead}$ & 5 & µm\\
        Lead width (small junct.) & $w_\text{lead,s}$ & 102 & nm\\
        Lead width (large junct.) & $w_\text{lead,l}$ & 192 & nm\\
        Bottom lead thickness & $t_\text{lead,btm}$ & 30 & nm\\
        Top lead thickness & $t_\text{lead,top}$ & 80 & nm\\
        Bandage size & $A_\text{band}$ & $3\times 3$ & µm$^2$\\
        Bandage thickness & $t_\text{band}$ & 300 & nm\\
        \bottomrule
    \end{tabular}
    \caption{Design parameters for Josephson junction leads and bandages dimensions.}
    \label{tab:target_junctions}
\end{table}

The niobium-based devices were fabricated on a high-resistivity ($>20\,\text{kOhm}$) (100)-silicon wafer as detailed in Ref.~\cite{ColaoZanuz2025}. 
Immediately prior to loading the samples for the deposition of the Manhattan-style~\cite{Potts2001, Kreikebaum2020} Josephson junctions, the samples were ashed at $200\,$W for $20\,$s at $1\,$mbar in a barrel asher. 
After depositing $30\,$nm of aluminum, the junction barrier was formed by oxidation for 20 minutes at $10\,$Torr in a $15/85\%$ $\text{O}_2$/Ar mixture. 
The second junction arm and aluminum bandages were fabricated using the same process steps as in Ref.~\cite{ColaoZanuz2025}. 
The normal-state resistance of the junctions was measured, but the junctions were not SEM-imaged for this process. 
We therefore list the targeted dimensions for the junctions and bandages in Tab.~\ref{tab:target_junctions}.

The tantalum thin films were prepared in a similar manner as discussed in Ref.~\cite{ColaoZanuz2025} on identical wafers.
Changes were the introduction of an additional baking step under vacuum prior to deposition of tantalum, to remove contaminants~\cite{Chayanun2024}. 
To achieve an $\alpha$-phase tantalum film, the wafer was heated to $600\,^\circ$C~\cite{Singer2024} and tantalum was DC-sputtered from a confocal two-inch target at a pressure of $2\times10^{-3}\,$mTorr and power of $200\,$W. 
The deposition rate was $7.5\,\text{nm}/\text{min}$. 
In total, we fabricated and measured five devices~--~A, B, C, D, and E~--~over the course of the experiments. 
Devices A, D, and E are niobium-based, while devices B and C are tantalum-based.

\section{Experimental Setup}
\label{app:setup}

\begin{figure*}
\includegraphics[width=\linewidth]{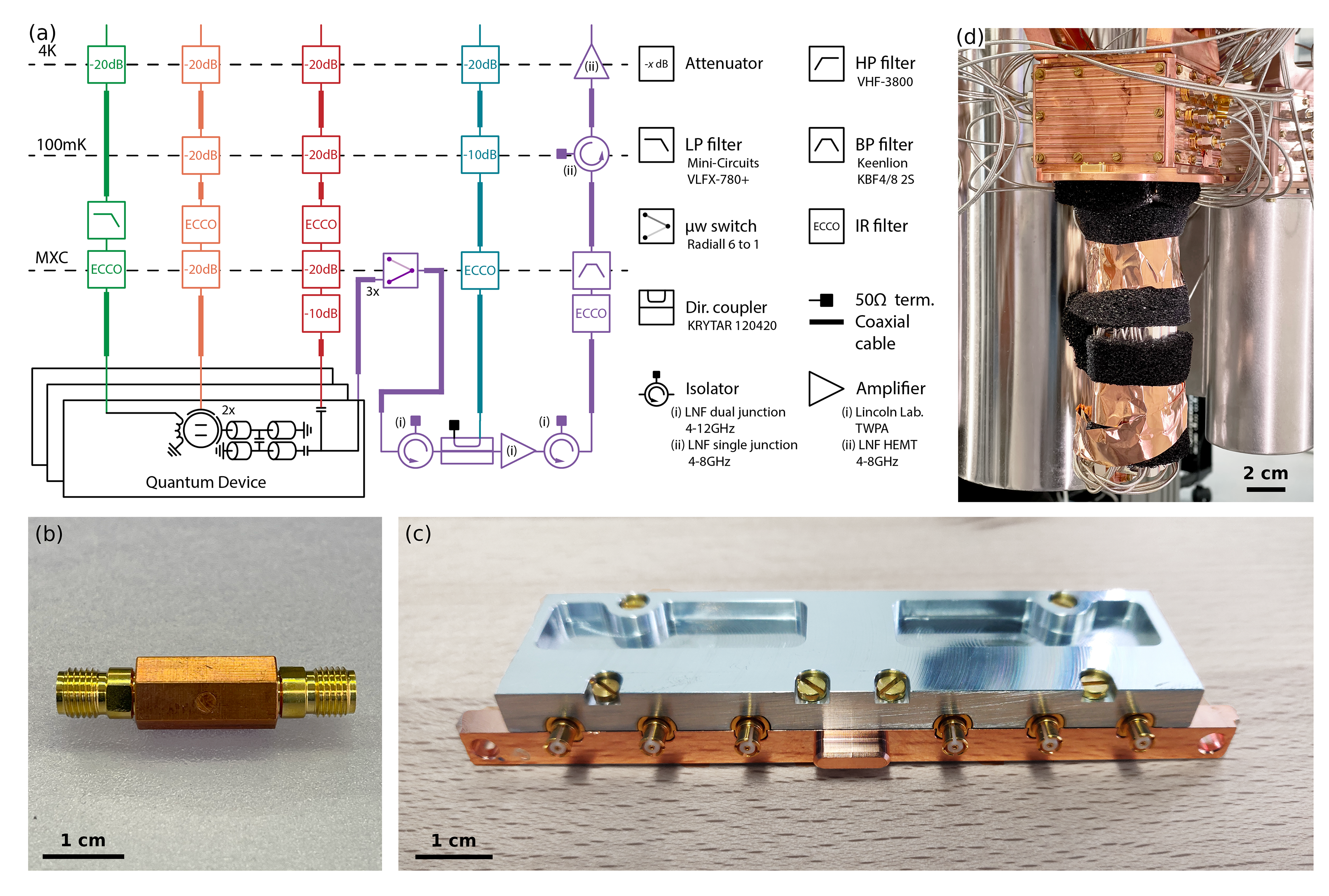}
\centering
\caption{Experimental setup. 
(a) Diagram of the cryogenic wiring for the setup. 
We show the line configuration for flux lines in green, for charge lines in orange, for readout input in red, for readout output in purple and for TWPA pump lines in teal. 
Components are thermalized at the indicated temperature stage. 
(b) Photograph of an Eccosorb filter for drive lines. 
(c) Photograph of the sample package consisting of a copper base and aluminum lid enclosing the quantum device. 
(d) Photograph of the sample mount with Eccosorb HR foam installed and removed shield assembly. 
The copper mezzanine together with the shield assembly (not shown in the image) provide a light-tight enclosure for the sample package.}
\label{fig:A}
\end{figure*}

Our experimental setup is similar to the one described in Ref.~\cite{Krinner2019}. 
We use stainless steel coaxial cables to route control and readout signals to the base plate of the cryostat, either by thermalizing the outer conductor of the cables using mechanical clamps or using a clamped attenuator at each temperature stage. 
For charge lines (single qubit drive and readout input) we use an attenuation of $60\,$dB in total and install an Eccosorb filter in each line, thermalized at the base plate of the cryostat. 
For the flux lines we use $20\,$dB of attenuation at the 4K stage, a low-pass filter and an Eccosorb filter at base, see Fig.~\ref{fig:A}a. 
Different to the setup described in Ref.~\cite{Krinner2019} we also install Eccosorb filters in the readout output lines, see Sec.~\ref{sec:filters}.
Furthermore, we make use of a Radiall Subminiature microwave switch mounted on the base plate, to multiplex a readout output line between devices. 
We note that the presence of the microwave switch does not appear to impact the measured quasiparticle tunneling rates, as evidenced by the comparable rates observed when infrared filters are placed before or after the switch (i.e., before and after the TWPA in Fig.~\ref{fig:filters}b).

The Eccosorb filters are made in-house and consist of a cylindrical copper enclosure filled with infrared absorbing Eccosorb material, housing a center conductor. 
We use a mixture of CR110 contained in a housing with inner diamter of $4.6\,$mm and a length of $10\,$mm for the charge and readout output lines, achieving a $3\,$dB cutoff frequency of $14\,$GHz and CR124 with a housing of the same inner diameter and a length of $2\,$mm for the flux lines, with a cutoff frequency of $4\,$GHz.
We show an image of one of the Eccosrob drive line infrared filters in Fig.~\ref{fig:A}b.

The device is glued to a copper base using GE varnish, wirebonded to a PCB, and covered by an aluminum lid (see Fig.~\ref{fig:A}c).
At the interface between the copper base and the aluminum lid, notches are added to improve light tightness. 
A potential entry point for infrared radiation remains at the seam around the microwave connectors of the sample package, which~--~although mechanically secured against a copper cold finger~--~may still allow leakage.
The sample package is placed into a shield assembly consisting of a copper shield for providing a well thermalized environment, an aluminum shield and two mu-metal shields to screen static magnetic fields, listed from the inner- to the outermost shield.
Signal lines are routed into the shield assembly via feedthrough connectors in the side plates of a light-tight copper mezzanine, see Fig.~\ref{fig:A}d. 
Additionally, for some experiments we add Eccosorb-HR foam~\cite{EccosorbHRWebsite} inside the copper shield, as shown in Fig.~\ref{fig:A}d. 
The foam is held in place using copper tape, and once the shield assembly is closed, it remains in compressive contact with the inner surface of the cylindrical copper shield, providing thermal coupling.
Despite the simplicity of this installation, we observe effective suppression of quasiparticle tunneling and no signs of insufficient thermalization of the foam.

\section{Quasiparticle Tunneling Measurement Procedure and Analysis} 
\label{app:procedure}

\begin{figure}[htbp!]
\includegraphics[width=\linewidth]{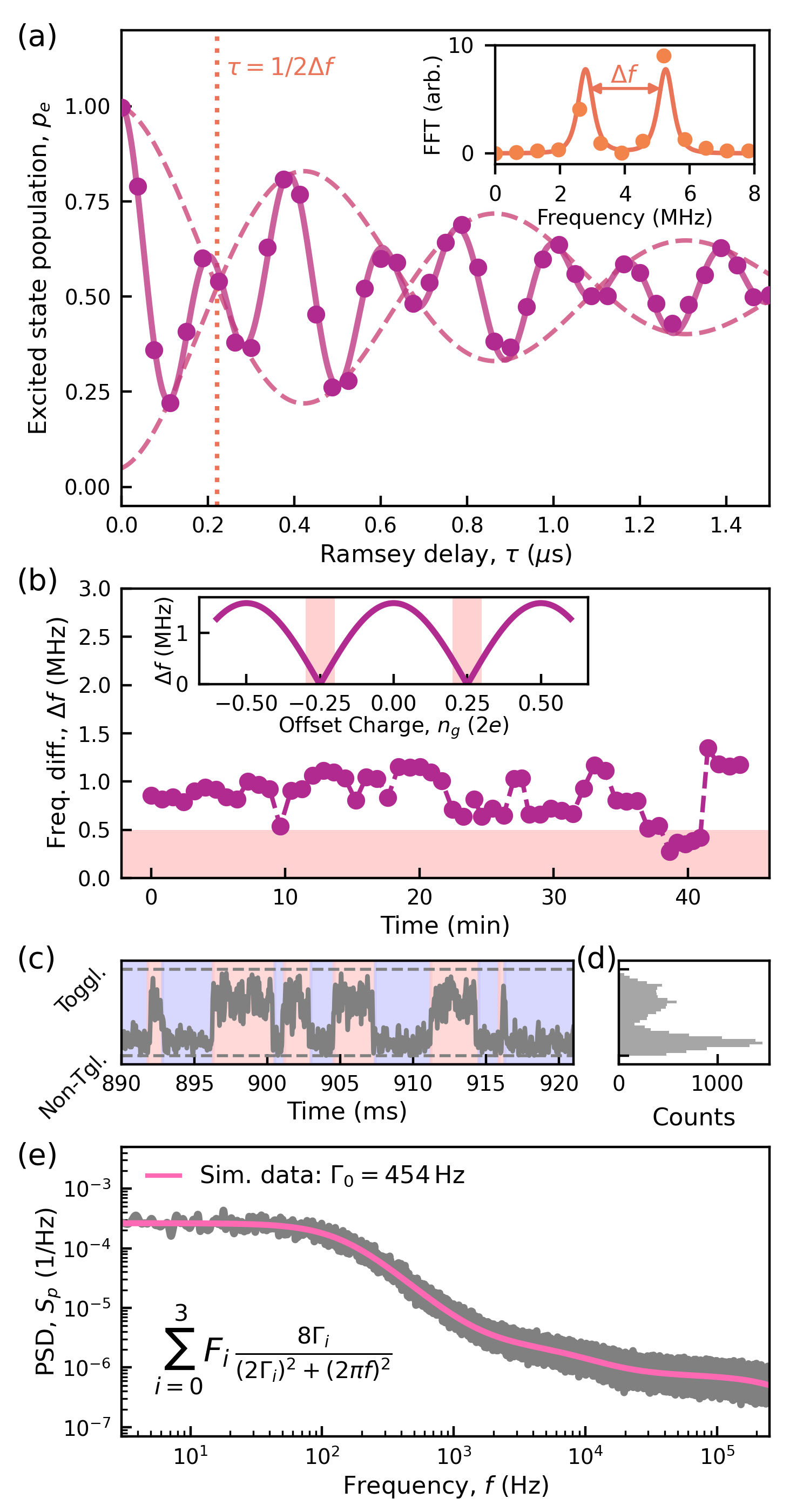}
\centering
\caption{Quasiparticle tunneling rate measurement protocol. 
(a) Averaged Ramsey experiment resolving beating in oscillation stemming from two frequency components. The data is fitted using two cosinusoidal frequency components and decaying exponentially. 
Inset: Fast Fourier transform of data in (a) fitted by two Lorentzians. 
(b) Time series of extracted frequency differences $\Delta f$ over a 40-minute interval. 
The region where $\Delta f \leq 0.5$\,MHz is shaded in red; see main text for details. 
Inset: Calculated frequency difference $\Delta f=|f_e-f_o|$ as a function of offset charge $n_g$ with region for which $\Delta f\leq0.5\,$MHz shaded in red.
(c)~Excerpt of simulated time trace of absolute differences $d_i$ (Gaussian average, $\sigma=10$) and (d)~corresponding histogram (30 bins) of tunneling protocol sampled from Markov chain model including gate and measurement infidelities. Color scheme as in Fig.~\ref{fig:irsweep}a.
(e) Power spectral density of simulated data. We fit the tunneling rate as described in the main text and extract a tunneling rate of $454\,$Hz.}
\label{fig:methods}
\end{figure}

The experimental protocol consists of two interleaved measurements: an averaged Ramsey experiment to determine the two frequency components $f_\mathrm{e}$ and $f_\mathrm{o}$, and the quasiparticle tunneling rate extraction experiment.
In the averaged Ramsey experiment~\cite{Krantz2019}, we intentionally detune the drive by $4\,$MHz from the center frequency and extract the two frequency components by fitting the data to a decaying cosine model with two frequency components, as illustrated in Fig.~\ref{fig:methods}a.
Since we do not actively control the qubit’s gate charge, the offset charge drifts on a timescale of minutes.
Consequently, the frequency difference $\Delta f = |f_\mathrm{e} - f_\mathrm{o}|$ varies between 0 and the maximum observed charge dispersion of approximately $5\,$MHz, and must be recalibrated prior to extracting the quasiparticle tunneling rate.

To extract the quasiparticle tunneling rate, we perform the experiment sequence introduced in Ref.~\cite{Riste2012a} and shown in Fig.~\ref{fig:device}b, driving the qubit at the intermediate frequency $(f_e+f_o)/2$. 
The wait time between the two Ramsey pulses in the quasiparticle tunneling rate extraction experiment is inversely proportional to $\Delta f$, and the experiment is susceptible to dephasing during the evolution along the equator of the Bloch sphere (see Fig.~\ref{fig:device}c,~d). 
As a result, the dephasing time $T_2^*$ imposes a lower bound on acceptable $\Delta f$ values. 
Given a typical $T_2^*$ of $10\,$µs, we conservatively set this lower limit to $\Delta f = 0.5\,$MHz, corresponding to a maximal wait time of $2\,$µs, which is well within the qubit’s coherence time.
We show the beating frequencies extracted over a 44-minute interval in Fig.~\ref{fig:methods}b, with the region $\Delta f \leq 0.5\,$MHz shaded in red.
Out of 58 measurements, five (minutes $39$-$41$), or roughly 9\%, yielded $\Delta f \leq 0.5$\,MHz, in reasonable agreement with the 19\% of offset charges for which the calculated frequency difference is below 0.5\,MHz (see inset of Fig.~\ref{fig:methods}b), when assuming a constant distribution of realized offset charges.
In these instances, as we do not actively control the qubit offset charge, we did not proceed with the rate extraction experiment but repeated the averaged Ramsey sequence until the frequency difference again exceeded 0.5\,MHz.

From the rate extraction experiment, we obtain a time trace of integrated single-shot in-phase and quadrature pairs, from which we determine a series of assigned qubit states $m_i\in\{0,1\}$ using the methods discussed in Ref.~\cite{Walter2017}.
In our restless implementation, we distinguish between two cases: the readout results toggling between $\ket{0}$ and $\ket{1}$ or remaining steady in either state. 
To analyze the data, we first compute the absolute value of the difference between subsequent measurement outcomes of the time trace of assigned qubit states $m_i$, i.e., computing $d_i=|m_{i+1}-m_i|$.
This yields a sequence $d_i$ that indicates whether the qubit is in a toggling $d_i=1$ or non-toggling state $d_i=0$ (see Fig.~\ref{fig:irsweep}a for a Gaussian-averaged time trace with a filter of width $\sigma=10$). 
We then apply a discrete Fourier transform to the unfiltered data with the frequencies set by the sampling rate and the duration of the sequence and compute the squared magnitude of the result to obtain the power spectral density. 
Since the sequence sampling rate depends on $\Delta f$, it can vary between different runs. 
To account for this, we interpolate the individual power spectral densities to a common frequency grid defined by the smallest frequency spacing and normalize each by dividing by the duration of the corresponding time trace before averaging across different runs.

To assess the impact of qubit coherence and measurement infidelities on our ability to extract quasiparticle tunneling rates, we construct a Markov process model~\cite{vanKampen2007} of the rate extraction experiment.
We simulate time traces using typical experimental parameters, assuming a tunneling rate of $460\,$Hz, a qubit relaxation time $T_1 = 20\,$µs, a dephasing time $T_2 = 10\,$µs, and a readout misclassification error of 2\%. 
Quasiparticle tunneling events, qubit decay, dephasing, and readout errors are sampled from Poisson distributions with expectation values determined by the corresponding rates.
In our Markov model, the qubit is assumed to be in one of four possible combinations of qubit state (ground or excited) and charge parity (even or odd).
The next state in the Markov chain is uniquely determined by the current state and by whether a tunneling, decay, dephasing, or readout error event occurs.

The resulting time trace exhibits a behavior similar to that observed in the experiment, most notably that the toggling state for even charge parity is not fully reached.
This can be seen from the Gaussian-averaged time trace of the absolute differences $d_i$ in Fig.~\ref{fig:methods}c, which does not attain the value of one (upper gray dashed line) corresponding to an ideal toggling signal.
We attribute this to systematics inherent to the restless method: while the non-toggling (odd) state is predominantly detected in the ground state due to energy relaxation, the toggling (even) state is more susceptible to errors from relaxation during readout.
Additionally, the experimental data shows a larger spread in the Gaussian-averaged time trace of the absolute differences $d_i$ for the even-parity state~--~a feature that is also reproduced, albeit less pronounced, in the simulated data, as indicated by the histogram in Fig.~\ref{fig:methods}d.

We extract the quasiparticle tunneling rate by fitting a sum of three Lorentzian functions to the power spectral density and identifying the tunneling rate as the center frequency of the lowest-frequency Lorentzian component. 
For the simulated dataset, this procedure yields a tunneling rate of $454\,$Hz, corresponding to a deviation of approximately $1$-$2\%$ relative to the value of $460\,$Hz used for the simulation (see above).
Similar deviations of a few percent are observed for simulated rates in the range of $100$ to $10\,000\,$Hz. 
Only for rates approaching the upper end of the experimentally measured range ($>100\,$kHz) do the extraction errors increase to approximately 20\%, which remains comparable to the spread in measured rates across nominally identical qubits.

In addition to the Lorentzian component associated with quasiparticle tunneling, we observe an additional Lorentzian at approximately $100\,$kHz in both the experimental datasets and the simulated power spectral density.
We attribute this feature to misclassification errors.
Some experimental datasets also show a Lorentzian component at approximately $10\,$kHz, whose frequency does not depend on experimental parameters, such as infrared filter configuration. 
This feature can be reproduced in the Markov process simulations by introducing variations in the single-qubit gate error rate.
We hypothesize that such variations may originate from ground noise coupling to the offset charge, thereby modulating gate fidelities.
See Fig.~\ref{fig:methods}e for a fit of the simulated dataset.

\section{Quasiparticle Density Scaling in the Presence of Black-Body Radiation}
\label{app:scaling}

Here, we discuss a simple model to estimate the quasiparticle density induced by infrared radiation from an electrically heated black-body source~\cite{Diamond2022}. The resulting steady-state quasiparticle density~--~and thus the quasiparticle tunneling rate~--~scales either quadratically or linearly with the electrical power emitted at the radiator, depending on which quasiparticle-loss mechanism dominates. However, this simplified model assumes that all emitted radiation contributes equally to quasiparticle generation, which may not be the case: only photons with energies above twice the superconducting gap can break Cooper pairs. At the radiator temperatures considered here, a significant fraction of the emitted power likely lies below this threshold and therefore does not contribute to quasiparticle generation.

The quasiparticle density $x_\mathrm{qp}$ in a superconductor in steady state is determined by a balance of the quasiparticle generation rate $g$, the single-particle  trapping rate $s$, and the two-particle recombination coefficient $r$~\cite{Lenander2011}: 

\begin{align}
\frac{dx_\mathrm{qp}}{dt} &= g - s x_\mathrm{qp} - r x_\mathrm{qp}^2 = 0.
\end{align}

\noindent Since recombination involves two quasiparticles, the corresponding term scales as $x_\mathrm{qp}^2$.
Both the coefficients $s$ and $r$ depend on material properties such as phonon-electron coupling or energy gaps~\cite{Catelani2011, Lenander2011}.
If trapping dominates ($s \gg rx_\mathrm{qp} $), the steady-state solution simplifies to

\begin{align}
x_\mathrm{qp} = \frac{g}{s},
\label{eq:steady1}
\end{align}

\noindent if recombination domintes ($rx_\mathrm{qp} \gg s$), it simplifies to

\begin{align}
x_\mathrm{qp} = \sqrt{\frac{g}{r}}.
\label{eq:steady2}
\end{align}

We assume that the dominant quasiparticle generation mechanism is the absorption of photons emitted by a black-body radiation source. The corresponding production rate $g$ thus depends on the temperature $T$ of the radiator. To model $T$, we consider a thermal balance in steady state:

\begin{align}
C \frac{dT}{dt} &= P - G(T - T_b) = 0,
\end{align}

\noindent where $C$ is the heat capacity of the radiator, $P$ is the applied electrical power, and $G$ is the thermal conductance to a bath at temperature $T_b$.
The heat transport is assumed to be governed by conduction rather than radiative cooling. 
According to the Wiedemann–Franz law~\cite{Jones1973}, the thermal conductance in this case is temperature dependent and can be written as

\begin{align}
G=\tilde{G}T,
\end{align}

\noindent where $\tilde{G}$ is a temperature-independent constant that depends on material and geometry.
Assuming $T_b \ll T$, we obtain

\begin{align}
T=\sqrt{\frac{P}{\tilde{G}}}.
\label{eq:radiator}
\end{align}

\noindent The thermal radiation power $P_\text{therm}$ emitted by a black body follows the Stefan–Boltzmann law:

\begin{align}
P_\text{therm}=\sigma AT^4,
\end{align}

\noindent where $\sigma$ is the Stefan-Boltzmann constant and $A$ is the surface area of the radiator at temperature $T$.
Inserting this expression, along with the previously derived temperature dependence of the radiator~\eqref{eq:radiator}, into the quasiparticle generation rate yields

\begin{align}
    g = \epsilon\sigma A\left(\frac{P}{\tilde{G}}\right)^2
\end{align}

\noindent where $\epsilon$ is an effective Cooper-pair-breaking efficiency constant. It quantifies the fraction of emitted black-body radiation that results in the breaking of Cooper pairs in the superconductor.
This efficiency is expected to depend on the superconducting energy gap (inversely), the reflectivity of the film, and the surrounding electromagnetic environment, including the presence of absorbing or reflective surfaces.
For simplicity, we neglect the spectral distribution of the emitted radiation and assume that most of the power lies above the energy gap.

Substituting the expression for $g$ into the steady-state solutions \eqref{eq:steady1} and \eqref{eq:steady2} for the quasiparticle density yields the following scaling relations:

\begin{align}
    x_\mathrm{qp}=\frac{\epsilon\sigma A}{s}\left(\frac{P}{\tilde{G}}\right)^2
\end{align}

\noindent for trapping-dominated scenarios and

\begin{align}
    x_\mathrm{qp}=\sqrt{\frac{\epsilon\sigma A}{r}}\frac{P}{\tilde{G}}
\end{align}

\noindent for recombination-dominated scenarios. Finally, we note that the quasiparticle tunneling rate is expected to be proportional to the quasiparticle density $x_\mathrm{qp}$~\cite{Catelani2012,Glazman2021}.

\begin{figure*}[t!]
\includegraphics[width=\linewidth]{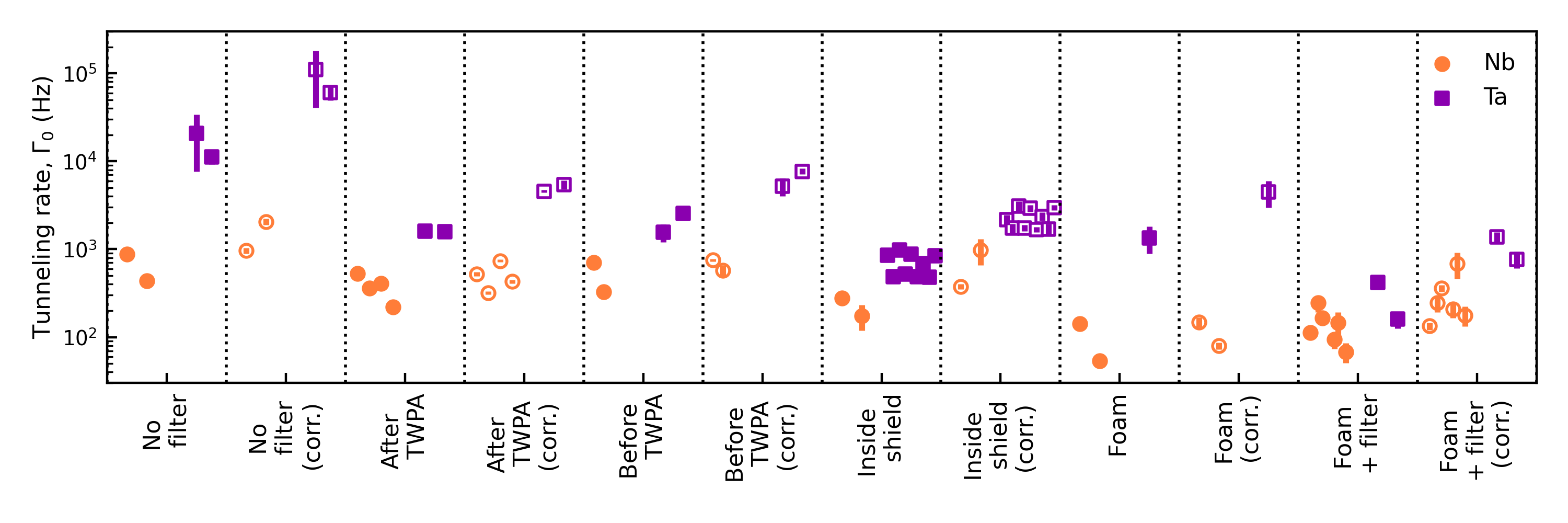}
\centering
\caption{Quasiparticle tunneling rates for niobium and tantalum qubits for different filter configurations. Open markers show data corrected for the experimentally observed time dependence, obtained by extrapolating the measurements from Fig.~\ref{fig:filters}b to a common reference time of $t = 1\,$day. Filled markers indicate the corresponding uncorrected values for comparison.}
\label{fig:filters_corr}
\end{figure*}

\begin{table*}[t]
    \centering
    \noindent
    \begin{minipage}[t]{0.5\textwidth}
        \vspace{0pt} % force top alignment
        \begin{tabular}{lp{2.2cm}cccc}
            \toprule
            Mat. & Configuration & $\Gamma_0$ (Hz) & Dev. & Qb. & $t$ (days) \\
            \midrule
            Nb & No filter & 883 $\pm$ 68 & A & 2 & 2.9 \\
            Nb & No filter & 438 $\pm$ 34 & D & 1 & 25.2 \\
            Ta & No filter & 20756 $\pm$ 13158 & B & 1 & 33.2 \\
            Ta & No filter & 11245 $\pm$ 2000 & B & 2 & 33.2 \\
            Nb & After TWPA & 531 $\pm$ 29 & A & 1 & 0.8 \\
            Nb & After TWPA & 358 $\pm$ 15 & A & 2 & 0.8 \\
            Nb & After TWPA & 407 $\pm$ 14 & A & 1 & 3.5 \\
            Nb & After TWPA & 219 $\pm$ 11 & A & 2 & 4.1 \\
            Ta & After TWPA & 1596 $\pm$ 57 & C & 1 & 8.9 \\
            Ta & After TWPA & 1583 $\pm$ 176 & C & 1 & 13.0 \\
            Nb & Before TWPA & 703 $\pm$ 26 & D & 1 & 2.2 \\
            Nb & Before TWPA & 325 $\pm$ 45 & E & 1 & 3.3 \\
            Ta & Before TWPA & 1566 $\pm$ 359 & B & 1 & 12.3 \\
            Ta & Before TWPA & 2537 $\pm$ 185 & C & 1 & 10.1 \\
            Nb & Inside shield & 279 $\pm$ 18 & A & 1 & 36.3 \\
            Nb & Inside shield & 175 $\pm$ 57 & A & 2 & 36.3 \\
            Ta & Inside shield & 858 $\pm$ 115 & B & 1 & 6.9 \\
            Ta & Inside shield & 488 $\pm$ 60 & B & 1 & 14.2 \\
            Ta & Inside shield & 982 $\pm$ 128 & B & 2 & 10.9 \\
            Ta & Inside shield & 523 $\pm$ 42 & B & 1 & 12.3 \\
            Ta & Inside shield & 875 $\pm$ 74 & B & 2 & 12.3 \\
            Ta & Inside shield & 488 $\pm$ 33 & B & 1 & 13.2 \\
            \bottomrule
        \end{tabular}
    \end{minipage}%
    \hfill
    \begin{minipage}[t]{0.5\textwidth}
        \vspace{0pt}
        \begin{tabular}{lp{2.2cm}cccc}
            \toprule
            Mat. & Configuration & $\Gamma_0$ (Hz) & Dev. & Qb. & $t$ (days) \\
            \midrule
            Ta & Inside shield & 686 $\pm$ 77 & B & 2 & 13.2 \\
            Ta & Inside shield & 479 $\pm$ 73 & B & 1 & 13.9 \\
            Ta & Inside shield & 838 $\pm$ 58 & B & 2 & 13.9 \\
            Nb & Foam & 141 $\pm$ 18 & A & 1 & 1.8 \\
            Nb & Foam & 54 $\pm$ 5 & A & 2 & 2.2 \\
            Ta & Foam & 1347 $\pm$ 456 & B & 2 & 12.2 \\
            Nb & Foam + Inside shield & 245 $\pm$ 52 & A & 1 & 1.0 \\
            Nb & Foam + Inside shield & 145 $\pm$ 48 & A & 2 & 25.9 \\
            Nb & Foam + After TWPA & 113 $\pm$ 11 & A & 1 & 7.4 \\
            Nb & Foam + After TWPA & 164 $\pm$ 15 & A & 1 & 5.1 \\
            Nb & Foam + After TWPA & 94 $\pm$ 20 & A & 2 & 5.3 \\
            Nb & Foam + After TWPA & 68 $\pm$ 17 & A & 2 & 7.4 \\
            Ta & Foam + Before TWPA & 417 $\pm$ 55 & C & 1 & 12.0 \\
            Ta & Foam + Before TWPA & 160 $\pm$ 34 & C & 1 & 26.6 \\
            \bottomrule
        \end{tabular}
    \end{minipage}
    
    \caption{Detailed description of the data set shown in Fig.~\ref{fig:filters}b. For each measurement, the material, filter configuration, tunneling rate $\Gamma_0$, label of the quantum device the qubits are from, qubit number, and time since cooldown $t$ are listed.}
    \label{tab:app_filters}
\end{table*}

\section{Filter Configuration Datasets}
\label{app:filter_data}

Our measurements of the quasiparticle tunneling rates acquired with different configurations of in-line filters and foam absorbers, see Fig.~\ref{fig:filters}b, were carried out at different times after the cooldown of the setup. 
Given that the tunneling rates decrease with time after initiating the cooldown, as discussed in Sec.~\ref{sec:decay}, we verify that our conclusions remain qualitatively valid despite the different times at which the measurements were taken.
We list the data for all individual measurements in Tab.~\ref{tab:app_filters}, including the time at which the measurement was taken, and the device and qubit used. 
Using the fitted time-dependence of the tunneling rates for niobium and tantalum, shown in Fig.~\ref{fig:decay}, we extrapolate corresponding rates at $t=1$\,day for all measurements.
In doing so, we assume that the time-dependence followed the same slope in all measurements for identical materials. 
Comparing adjusted and original data, our qualitative observations of the impact of different filtering strategies remain unchanged, see~Fig.~\ref{fig:filters_corr}.

\bibliographystyle{apsrev4-2-title-etal}
\bibliography{QudevRefDB, AdditionalRefs}

\end{document}